\documentclass[prl,twocolumn,aps,superscriptaddress,showpacs]{revtex4}
\usepackage{amsmath}
\usepackage{epsfig}
\usepackage[latin1]{inputenc}

\begin{document}

\title{Crystallization and melting of bacteria colonies and Brownian Bugs}

\author{Francisco Ramos} 
\affiliation{
Departamento de Electromagnetismo y F\'\i sica de la Materia and
Instituto de F\'\i sica Te\'orica y Computacional Carlos I, Facultad
de Ciencias, Univ. de Granada, 18071 Granada, Spain.  }

\author{Crist\'obal L\'opez}
\affiliation{Instituto de F\'\i sica Interdisciplinar y Sistemas Complejos IFISC (CSIC-UIB). Campus de la
Universidad de las Islas Baleares, E-07122 Palma de Mallorca, Spain.}

\author{Emilio Hern\'andez-Garc\'\i a}
\affiliation{Instituto
de F\'\i sica Interdisciplinar y Sistemas Complejos IFISC (CSIC-UIB). Campus de la
Universidad de las Islas Baleares, E-07122 Palma de Mallorca, Spain.}

\author{Miguel A. Mu\~noz} 
\affiliation{
Departamento de Electromagnetismo y F\'\i sica de la Materia and
Instituto de F\'\i sica Te\'orica y Computacional Carlos I, Facultad
de Ciencias, Univ. de Granada, 18071 Granada, Spain.  }
\date{\today}

\begin{abstract}
  Motivated by the existence of remarkably ordered cluster arrays of
  bacteria colonies growing in Petri dishes and related problems, we
  study the spontaneous emergence of clustering and patterns in a
  simple nonequilibrium system: the individual-based interacting
  Brownian bug model. We map this discrete model into a continuous
  Langevin equation which is the starting point for our extensive
  numerical analyses. For the two-dimensional case we report on the
  spontaneous generation of localized clusters of activity as well as
  a melting/freezing transition from a disordered or isotropic phase
  to an ordered one characterized by hexagonal patterns. We study in
  detail the analogies and differences with the well-established
  Kosterlitz-Thouless-Halperin-Nelson-Young theory of equilibrium
  melting, as well as with another competing theory. For that, we
  study translational and orientational correlations and perform a
  careful defect analysis. We find a non standard one-stage,
  defect-mediated, transition whose nature is only partially
  elucidated.
\end{abstract}
\pacs{
02.50.Ey, 
05.10.Gg, 
05.40.-a, 
87.18.Ed  
}
\maketitle

\section{Introduction}
Spontaneous emergence of clustering is a widespread phenomenon in
population biology, ecology, material science, and other fields
\cite{Murray,Levin}.  Either passive or active ``particles'' (trees,
bacteria, plankton, etc.) bunch together forming dense localized
clusters embedded in an, otherwise, almost empty surrounding space.
The patchy distribution of plankton in the ocean surface
\cite{plankton}, the spatial distribution of vegetation in semi-arid
regions \cite{vegetation}, or the fascinating patterns generated by
bacteria grown in Petri dishes \cite{Budrene,BJ} are some examples.
Some simple mechanisms leading to clustering have been described:

(i) Non-interacting inertial particles moving in a fluctuating environment (as
a turbulent flow)
may become clustered \cite{Coalescence}.
This \textit{path coalescence} mechanism has been claimed to
contribute to plankton patchiness.

(ii) Branching and annihilating Brownian particles (also called
\textit{super-Brownian processes}) 
tend to bunch together. This type of clustering has its roots in the
fact that offsprings are created within a local neighborhood of their
parents and die anywhere, giving rise to an overall tendency to form
localized colonies \cite{Young}.

(iii) In some cases, clustering stems from the interaction with a
second ``field'' influencing the dynamics (water in models for
vegetation patchiness \cite{vegetation}, nutrients or chemicals
in models for chemotactic bacteria clustering \cite{Levin}, etc.)
through some type of feedback mechanism.
Reaction-diffusion equations have proven to be a
convenient tool to model clustering at a mesoscopic level in this case
\cite{Murray,Cross,Walgraef,Fuentes}.  

Contrarily to the cases above, there are clustering models in which
individuals interact with each other in a direct way. For example,
their effective interaction can be such that the reproduction rate (or
some other individual dynamical property) is diminished by a factor
depending on the relative abundance of neighboring individuals.  As a
well documented example of this, let us mention the empirically
observed {\it Janzen-Connell} effect in ecology: owing to various
circumstances (presence of specialized pests or predators, competence
for resources, etc.), the effective reproduction rate of an individual
decreases with the number of conspecific specimens surrounding it
\cite{JC}. Similar effects appear also in bacteria colonies, social
phenomena, and in systems exhibiting collective motion
\cite{collective}.

In many of these systems, clusters are distributed in space in a
disordered way but, very remarkably, in some other cases they {\it
  self-organize} forming rather regular {\it patterns}. As a simple
example to bear in mind, let us focus on the strikingly ordered
patterns self-generated by {\it Escherichia Coli, Salmonella
  typhimorium}, and other bacteria grown in Petri dishes. When grown
in a substrate of nutrients under adequate conditions these
bacteria colonies structure themselves into clusters which, in their
turn, self-organize in spirals, squares, or crystal-like hexagonal
arrays (as a beautiful illustration, see Fig.1 in \cite{Budrene} or
page 529 of \cite{BJ}).

How ordered can such two-dimensional patterns be?  This question
resembles very much the problem of crystal ordering in equilibrium
solids and the existence of a melting (freezing) solid/liquid
(liquid/solid) transition. For two-dimensional systems in
thermodynamic equilibrium the Mermin-Wagner-Hohenberg (MWH) theorem
\cite{MW} rules out continuous symmetries to be spontaneously broken
in the presence of fluctuations.  This covers the case of
translational invariance and, therefore, two dimensional
crystals/solids cannot exhibit true long-range translational order
(destroyed by low-energy Goldstone modes). Indeed, the most popular
equilibrium melting theory (to be described in some detail below
\cite{KT,KTHNY,Strandburg,Nelson}) predicts the melting transition to
occur in two-stages, and to include an intermediate {\it hexatic}
phase in between a {\it quasi long-range ordered} solid phase and the
disordered (or isotropic or liquid) phase.  Alternatively, a competing
theory \cite{Chui} predicts a unique discontinuous transition between
such two phases.

Note, notwithstanding, that all the clustering problems described
above occur away from thermodynamic equilibrium.  Actually, many of
them exhibit absorbing states, a quintessential example of
irreversible nonequilibrium dynamics (fluctuations can lead from
activity to the absorbing state, but not the other way around
\cite{AS}).

Does the MWH theorem applies to nonequilibrium problems as
self-organized bacteria colonies? Do they exhibit a melting/freezing
transition similar to that occurring in equilibrium? Do they have an
hexatic phase?

Nonequilibrium problems violating the MWH theorem are well known. A
notorious example is provided by the model of self-propelled particles
proposed by Vicsek {\it et al.} \cite{collective} which illustrates
how communities (flocks, schools) of interacting individuals (birds,
fishes) move coherently, with {\it true long-range order}, adopting a
collective non-vanishing velocity (which breaks the continuous
rotational symmetry). Soon after the introduction of such a model it
was proved in an elegant way that, indeed, true long-range order
emerges owing to inherently nonequilibrium effects \cite{TT} (see also
\cite{Racz} for a related, though different, nonequilibrium mechanism
leading to true long-range two-dimensional ordering). Is the MWH
violated in a similar way for the clustering problems described above?

Aimed at answering the previously posed questions and to scrutinize the role
of fluctuations in nonequilibrium clustered systems exhibiting self-organized
patterns, we analyze here a simple model of individual interacting particles
\cite{bugs1,bugs2,framos} exhibiting clustering as well as ordered patterns.

Our main conclusions are as follows: while we find a solid phase with quasi
long-range translational order and a disordered phase in analogy with
equilibrium situations, we do not find any evidence either of the hexatic
phase predicted by the most popular theory for equilibrium melting
\cite{KT,KTHNY,Strandburg,Nelson}, nor we find the discontinuous transition
predicted by competing theories \cite{Chui}. Instead, we report on a {\it
  one-stage continuous transition} analogous to what was reported in numerical
investigations of other equilibrium \cite{JJ} and nonequilibrium \cite{QG}
melting problems. An explanation justifying our findings, supported by the
analysis of topological defects and likely to be valid for other systems, is
proposed.

The paper is structured as follows. First we define the model,
describe its basic ingredients and introduce a Langevin equation
capturing all its relevant traits at a continuous level. We discuss
briefly its main phenomenology: emergence of clusters and ordering
into hexagonal patterns.  Afterwards, we report on its numerical
analysis paying attention to the melting transition (including a
careful analysis of topological defects) and compare our results with
standard equilibrium melting theories.  Finally, we discuss our
findings from a general perspective and present the conclusions.

\section{The Interacting  Brownian bug model and its Langevin representation}

The {\it individual-based} model we study here, the ``interacting Brownian
bug'' (IBB) model, was recently introduced by two of us \cite{bugs1,bugs2}.
It consists of branching-annihilating Brownian particles (bugs, bacteria)
which interact with each other within a finite distance, $R$ \cite{bugs1}.
Particles move off-lattice in a $d$-dimensional $[0,L]^d$ interval with
periodic boundary conditions, obeying a dynamics by which particles
can:
\begin{itemize}
\item {\it diffuse} (at rate $1$) performing Gaussian random jumps of
  variance $2D$,

\item {\it dissappear spontaneously} (at rate $\beta_0$),

\item {\it branch}, creating an offspring at their same spatial coordinates
 with a density-dependent rate $\lambda$:
\begin{equation}
\lambda(j)= max  \{ 0, \lambda_0- {N_R(j)}/{N_s} \},
\label{raterep}
\end{equation}
where $j$ is the particle label, $\lambda_0$ (reproduction rate in
isolation) and $N_s$ (saturation number) are fixed parameters, and
$N_R(j)$ stands for the number of particles within a radius $R$ from
$j$.
\end{itemize}
The control parameter is $\mu=\lambda_0-\beta_0$ while
 the function $max()$ enforces the transition rates positivity.
See \cite{otros} where similar models in which the death or the
diffusion rate are density-dependent are studied.

The main phenomenology of the IBB is as follows
\cite{bugs1,bugs2,framos}. For large values of $\mu$ there is a
stationary finite density of bugs (active phase) while for small
values the system falls ineluctably into the vacuum (absorbing phase).
Separating these two regimes there is a critical point at some value
$\mu_c$, belonging to the directed percolation (DP) universality class
\cite{framos} characterizing in a robust way transitions into
absorbing states \cite{AS}.

\begin{figure}
\begin{center}
\epsfig{figure=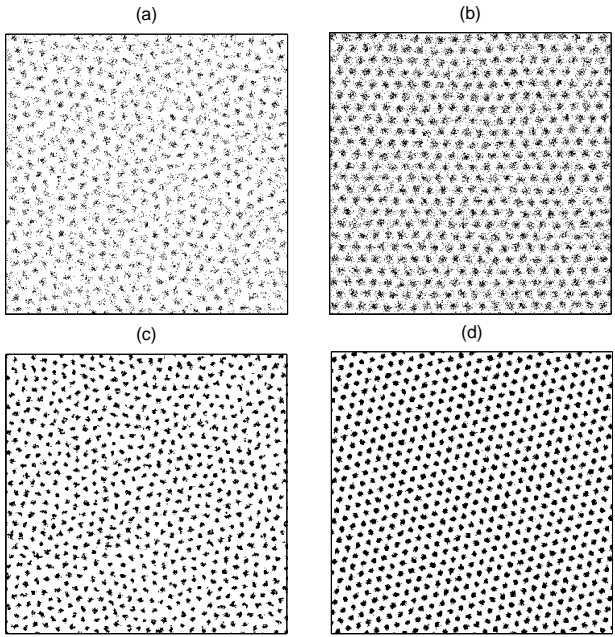,width=\columnwidth,angle=0}
\end{center}
\caption{Upper panels: Snapshots of two-dimensional IBB model in its
  stationary state (time $t=2 \times 10^5$ Monte Carlo steps) in the:
  (a) disordered active phase, $\mu=1.0$, and (b) the
  ordered/patterned active one $\mu=2.0$; other parameter values are:
  $N_s=50$, $R=0.1$, $D=10^{-5}$.  Lower panels: Snapshots of the
  two-dimensional Langevin equation, Eq.(\ref{langevin}) in its
  stationary state for $\mu=2.0$, $D=0.25$, $R=10$, $N_s=50$ in the:
  (c) disordered active phase $\sigma=2.5$ and (d) the
  ordered/patterned active one, $\sigma=1.5$. Observe that the 2
  snapshots to the right (b and d) have a crystal-like hexagonal
  ordering, absent in the other two (a and c). Also, the crystal-like
  packing is more evident in the continuous model for the chosen
  parameters, but no qualitative difference exist between the upper
  and the lower panels.}
\label{Snapshots}
\end{figure}

In the active phase, owing to the local-density dependent dynamical
rules particles group together forming clusters (see Fig.
\ref{Snapshots} (a) and (b)) provided that the diffusion constant is
small enough (for large values of D, homogeneous distributions are
obtained). Such clusters have a well-defined typical diameter (which
can be analytically estimated \cite{nb}) and a characteristic number
of particles within, which depend on the parameters $R$, $N_s$, and
$\mu$ \cite{bugs1,bugs2}.  Well inside the active phase, when the
clusters start filling the available space they {\it self-organize} in
spatial structures with remarkable hexagonal order (see
Fig.\ref{Snapshots} (b)).

From the theoretical side, an important breakthrough is that the IBB
model can be cast into a continuous stochastic equation \cite{bugs1}.
Indeed, by applying standard Fock-space (Doi-Peliti \cite{GF})
techniques, a Langevin equation including the main relevant traits of
the problem in a parsimonious way can be derived (see Appendix A in
\cite{bugs1} and references therein). The Langevin equation for the
local density \cite{density} of bugs $\rho({\bf x},t)$ reads (in the Ito
representation)
\begin{eqnarray}
\dfrac{\partial \rho(\textbf{x},t)}{\partial t} &=& \mu 
\rho(\textbf{x},t)+D \nabla^{2} \rho(\textbf{x},t)
\label{langevin}
 \\ \nonumber
&-&\frac{\rho(\textbf{x},t)}{N_s}\int_{\vert \textbf{x}-\textbf{y}
\vert < R} d \textbf{y} \rho(\textbf{y},t)+\sigma
\sqrt{\rho(\textbf{x},t)}\eta(\textbf{x},t),\\ \nonumber
\end{eqnarray}
where the noise amplitude $\sigma$ is a function of the microscopic
parameters and $\eta(\textbf{x},t)$ is a normalized Gaussian white
noise.
This includes only the leading terms in a density expansion; for
instance, a higher order noise term appear in the mapping
\cite{bugs1} but it does not alter the results reported in what  follows 
in any significant way.

Note that, leaving aside the {\it non-local} saturation term, this
equation coincides with the Reggeon-field theory or Gribov process,
describing at a coarse-grained level systems with absorbing states in
the directed percolation (DP) class (see \cite{AS} and references
therein). Let us underline the presence of a square-root
multiplicative noise.  Note also that the deterministic part of
Eq.(\ref{langevin}), including a non-local saturation term, is
identical to the equation proposed by Fuentes {\it et al.}
\cite{Fuentes}.  Eq.(\ref{langevin}) is therefore a simple (the
simplest) stochastic generalization of such model.

The main advantage of the continuous Langevin equation above is that it allows
for analytical studies and permits to scrutinize the effect of fluctuations
(by just tuning the noise amplitude). It also constitutes a more general,
elegant, and compact formulation of the problem. For these reasons, we choose
Eq.(\ref{langevin}) as the starting point of our forthcoming analyses.

\section{Integration of the Langevin equation and first analyses}

Analytical studies of the deterministic part of Eq.(\ref{langevin})
(i.e. mean field analyses) have already been performed in
\cite{bugs1,bugs2,Fuentes}.  They permit, for instance, to understand
the wave-length instability mechanism leading to pattern formation, as
well as many other aspects. In order to analyze the full stochastic
Eq.(\ref{langevin}) and, in particular, its associated
melting/freezing transition, one needs to resort to computational
studies.

However, integrating numerically Langevin equations with square-root
noise is a highly non-trivial task; owing to the fact that for small
density values the square-root term (with multiplies a random number)
becomes larger in amplitude than the deterministic terms, standard
integration schemes (Euler, Runge-Kutta, etc.) lead ineluctably to
unphysical negative values of the density field \cite{dickman}, and
this pathology is not easily cured in any na\"ive way. Luckily, a very
efficient scheme, specifically devised to overcome such a difficulty,
has been recently proposed \cite{Dornic}.

The method is a {\it split-step} algorithm in which the system is
discretized in space and two evolution operations are performed at
each discrete time-step: (i) first, the noise term is treated in an
exact way, by sampling the conditional probability distribution coming
out of the (exactly solvable) associated Fokker-Planck equation at
each site. By sampling in an exact way such a distribution an output
is produced at each site. (ii) Afterwards, the remaining deterministic
terms are integrated using any standard scheme taking as the input at
each site the output of the previous step at each site.  More details
and applications can be found in \cite{Dornic}.

To implement the split-step scheme to integrate Eq.(\ref{langevin}) in
two dimensions, we discretize the space, by introducing a lattice of
linear size $L=256$ or $~512$ ($L \approx 1024$ is already at the
limit of our present computational power). We fix the discrete time
step to $\Delta t=0.25$, $R=10$, $N_s=50$, $D=0.25$ (which is small
enough to have clustering) and use either $\mu$ or $\sigma$ as a
control parameter. For all of the simulations reported here we
initialize the system with a homogeneous initial density,
$\rho({\bf{x}},t=0) = \rho_0$, leave it relax towards its
stationary state (reached typically after $10^ 5$ Monte Carlo steps),
and perform steady-state measurements (averaging over, at least,
$10^5$ configurations).

First of all, we verify that Eq.(\ref{langevin}) reproduces
qualitatively all the basic phenomenology of the microscopic IBB
model: (i) an absorbing phase for small values of $\mu$, (ii) an
active disordered phase, encountered by increasing $\mu$ for a fixed
$\sigma$ and (iii) an active ordered crystal-like phase which is
reached by further increasing the value of $\mu$ or alternatively,
fixing $\mu$ and reducing the noise amplitude, $\sigma$ (see
Fig.(\ref{Snapshots} (c) and (d))).

Separating (i) and (ii) there is a directed-percolation-like phase
transition, while our focus here is on the transition from the
disordered active phase (ii) to the ordered self-organized one (iii).
In order to study the effect of the noise on such a transition, from
now on, we fix $\mu=2.0$, well into the active phase, and use the
noise amplitude, $\sigma$, as a tuning parameter.  Fig.
\ref{Snapshots} (plots (c) and (d)) shows two snapshots obtained for
$\mu=2.0$ with $\sigma=2.5$ and $\sigma=1.5$ respectively.  While in
both cases clusters of localized activity ($\rho(\textbf{x}) \neq 0$)
exist, only in the second one clusters are self-organized into an
ordered hexagonal array.

\subsection{Cluster analyses}

A preliminary step towards a systematic cluster analysis is to have an
efficient method to detect and label them. We have implemented an algorithm,
based in the Hoshen-Kopelman one \cite{HK} as follows. First, to avoid
spurious clusters we apply a smoothening filter to the noisy field
$\rho(\textbf{x},t)$ in our simulations, which removes its short wavelength
fluctuations:
\begin{equation}
  \rho_{p}({\bf{x}},t)=\dfrac{1}{\pi R_p^2}\int_{|{\bf y}|=0}^{|{\bf y}|=R_p}\rho({\bf x}+{ \bf y},t),
  d{\bf{y}}
\end{equation} 
where $R_p$ is the cluster average radius (almost constant for the
considered parameter range). Then we define a ``smoothened'' binary
field, $\rho_p({\bf x},t)$ taking a value $0$ ($1$) wherever
$\rho_{p}({\bf x})<\Theta$ ($\rho_{p}({\bf{x}})>\Theta$). The threshold
$\Theta$ is fixed to an optimal value $\Theta =0.55$, found by trial
and error, and we verify that an excellent cluster identification (as
compared to visual inspection) is obtained.  The output is not very
sensible to the threshold choice although small variations can be
observed. Once the binary discretized field has been constructed, a
standard Hoshen-Kopelman algorithm is straightforwardly employed. It
generates a list of clusters, together with the list of spatial
coordinates ascribed to (the center of mass of) each of them.

Having described the main computational techniques, we now report on
various observables characterizing collective properties of clusters.
\begin{figure}
\begin{center}
\epsfig{figure=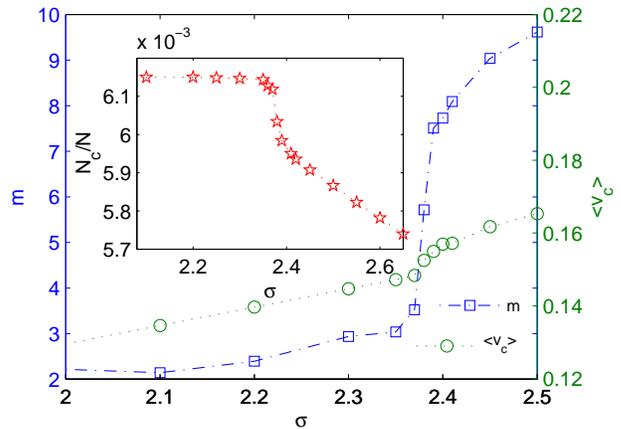,width=\columnwidth,angle=0}
\end{center}
\caption{ Main: cluster mobility per unit time (boxes and left axis) and mean
  cluster center of mass velocity (circles and right axis) as a function of
  $\sigma$ for $L=256$. Inset: Number of clusters per surface unit versus
  $\sigma$, for $L=256$.  }
\label{clusters1}
\end{figure}

The average cluster mobility, $m$, is defined as the standard deviation
of the excursion of the cluster center of mass (during a fixed-length
time interval), averaged over many different clusters in the steady
state
\begin{equation}
m  = \sqrt{\left\langle \left( 
 x_c-\left\langle x_c\right\rangle\right)^2 \right\rangle +
 \left\langle \left(y_c-\left\langle y_c\right\rangle\right)^2
 \right\rangle },
\end{equation}
where $(x_c,y_c)$ are the center of mass cluster coordinates and $\left\langle
  \cdot \right\rangle$ stands for averages in the steady state.

In Fig. \ref{clusters1} we plot $m$ as a function of the noise
amplitude $\sigma$; the mobility exhibits a sharp transition (sharper
upon enlarging the system size) from the low-noise phase in which the
clusters are almost frozen and localized in space to the high noise
one in which they move more freely.  The change of behavior occurs
around $\sigma \approx 2.39$.

Fig. \ref{clusters1} shows also the average velocity (in modulus) of
the cluster center of mass $(v_{x_c},v_{y_c})$
versus $\sigma$. While for small values of $\sigma$ we observe a
linear dependence between $\left\langle v_c\right\rangle$ and $\sigma$
\cite{linear}, this linear dependence is broken above certain noise
threshold (again around $\sigma \approx 2.39$), at which its
derivative exhibits a discontinuity.  Above the transition point the
velocity increases non-linearly with $\sigma$.

We have also measured (Fig.\ref{clusters1} inset) the number of
clusters per surface unit as a function of $\sigma$. While in the
disordered phase the density of clusters increases as the noise
strength is reduced, it remains constant (at a value corresponding to
the maximum capacity) once the threshold for an ordered structure is
reached.

The three described observables provide evidence of a melting/freezing
transition.  The change of behavior occurs in all cases at a unique
point, somewhere around $\sigma \approx 2.39$.  A more detailed finite
size scaling analysis would be required to pin down the critical point
with more accuracy using these observables.

The picture that emerges from these measurements is that clusters
emerge at a mesoscopic scale out of the nonequilibrium microscopic
rules and then, upon reducing the noise-amplitude, they self-organize
into frozen patterns with reduced mobility and velocity and with a
more compact packing. In this sense, clusters become the equivalent of
``particles'' in standard liquid/solid transitions.

\subsection{Structure Function Analysis.}
In order to obtain an alternative, more direct, estimation of the
location of the freezing/melting transition not relying on the
(computationally costly) identification of clusters, we analyze a
properly defined structure function. As the overall density varies
upon changing parameters and system size, it is convenient to define a
{\it normalized} version of the structure function as follows
\begin{equation}
S\left( k\right)=\left\langle \left| 
\textsl{F}(   {\bf k}  ) \right|^{2}\right\rangle_{k},
\end{equation} 
where
$F$ is the Fourier transform of the normalized density
\begin{equation}
\textsl{F}({\bf k})=\int_{L^d} \rho_{norm}({\bf x}) e^{-i {\bf k} \cdot {\bf x} } {\bf d x},
\end{equation}
${\bf k}$ is the momentum, $\left\langle \cdot \right\rangle_{k}$
stands for spherical averages over all two-dimensional vectors with
module $k=\vert \textbf{k}\vert$, and the normalized density
$\rho_{norm}$ is
\begin{equation}
\rho_{norm}\left({\bf x}\right)=\frac{\rho\left({\bf x}\right)}{\sqrt{\int_{L^d}\left|
 \rho\left({\bf x}\right) \right|^{2}}}.
\end{equation}
Using this normalized density, and by virtue of the Parseval's
identity,
$
\int_{L^{-d}} \left| \textsl{F}\left({\bf k}\right) \right|^{2} =
\int_{L^d}\left| \rho_{norm}\left({\bf x}\right) \right|^{2}=1,
$
it is guaranteed that $S\left( k\right)$ is normalized to unity for
all parameter values and system-sizes.

\begin{figure}
\begin{center}
\epsfig{figure=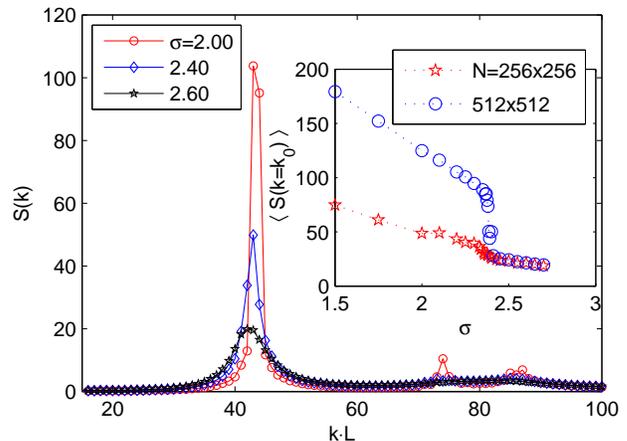,width=\columnwidth,angle=0}
\end{center}
\caption{Main plot: Structure function, $S(k)$, for three different values of
  the noise amplitude $\sigma$ and $L=512$. The peak around $k_0 \approx 42$
  becomes more pronounced as we go deeper into the ordered phase. Inset:
  $S(k_0)$ versus $\sigma$ for $L=256$ and $L=512$.}
\label{Structure}
\end{figure}

Fig.\ref{Structure} shows the structure-function for a size $L=512$ as
a function of $k$, for three different noise amplitudes:
$\sigma=2.20,~2.40,$ and $2.60$.  The three curves exhibits a very
pronounced first Bragg-peak at positions around $k_0 L \approx 43$
(which corresponds to $k_0 \approx 0.083(9)$ and therefore a
separation between clusters $R_0 \approx 11.9(1)$. Their respective
heights decrease with noise strength: the larger the noise the less
ordered the structure.  Nevertheless, perfect ordering (delta-peak at
a given mode) will never be reached as cluster internal fluctuations
enforce some dispersion in the structure function.

Given that $S(k)$ has been normalized, the height at $k_0$, $S(k_0)$,
constitutes a good measure of the degree of order. Actually, an abrupt
transition is obtained at about $\sigma \approx 2.39$ for the largest
available size (see (Fig.\ref{Structure})): above the transition
$S(k_0)$ decreases abruptly corresponding to the breakdown (melting)
of the ordered patterns. Also, observe in the inset of
Fig.\ref{Structure} that the degree of order is enhanced upon
enlarging the system size from $L=256$ to $L=512$. We will come back
to analyze this issue later.

\section{Two-dimensional melting theories: ~~~~~~~~~~~ a short review}

The evidence accumulated so far, both from cluster analyses and
structure function measurements, reveals the existence of a
melting/freezing transition, somewhere around $\sigma=2.39$. The key
question we face now is: does such a nonequilibrium transition exhibit
the well-known universal features of standard equilibrium
melting/freezing in two-dimensional systems?

Note, in addition, that here the number of clusters (particles) is not
constant but fluctuates, and its average value varies as a function of
the control parameter, $\sigma$. Is this relevant for the properties
at the melting/freezing transition?

Before tackling these problems, for the sake of completeness, and for
future reference, in this section we briefly summarize the main
results of the celebrated standard theory of two-dimensional melting:
the Kosterlitz-Thouless-Halperin-Nelson-Young (KTHNY) theory
\cite{KT,KTHNY,Strandburg,Nelson}. We also discuss briefly an
alternative competing theory.

The KTHNY is based on a statistical physics analysis of topological
defects, i.e. particles with a number of nearest neighbors (assuming a
Voronoi or Wigner-Seitz construction) other than $6$.  {\it
  Dislocations} perturb translational order and {\it disclinations}
hinder orientational order \cite{Nelson,Strandburg}.  A detailed
inspection of hexagonal ordering in the presence of fluctuations
reveals that disclinations correspond to free {\it monopoles}, either
five-fold or seven-fold, where five and seven refer to the number of
nearest neighbors as measured in a Voronoi or Wigner-Seitz
construction.  Analogously, dislocations can be identified with tight
pairs (i.e. {\it dipoles}) of a five-fold and a seven-fold
disclinations.  The disordered (or isotropic or liquid) phase is
characterized by the proliferation of defects: both monopoles and
dipoles.

The main prediction of the KTHNY theory is that, contrarily to what
happens in higher dimensional systems, where the melting occurs
discontinuously at a unique transition point, in two-dimensional
systems melting occurs in two stages. Translational and orientational
order lose their stability at different Kosterlitz-Thouless-like
\cite{KT} critical points where dislocations and disclinations,
respectively, unbind.

The theory assumes that in the solid phase there are nor free dipoles
nor monopoles, but only quadrupoles (low-energy excitations), that the
number of dislocations/dipoles throughout the first (melting)
transition is small and that they are generated progressively in a
smooth way as the temperature is risen. This allows to treat the
system at the melting transition as a weakly interacting gas of
dipoles/dislocations. An analogous assumption is made for
monopoles/disclinations at the second transition point where monopoles
unbind from dipoles.

The three phases put forward by the KTHNY theory are as follows
\cite{KTHNY,Strandburg,Nelson} (see Fig. \ref{KTHNY} for a graphical
illustration):
\begin{figure}
\begin{center}
\epsfig{figure=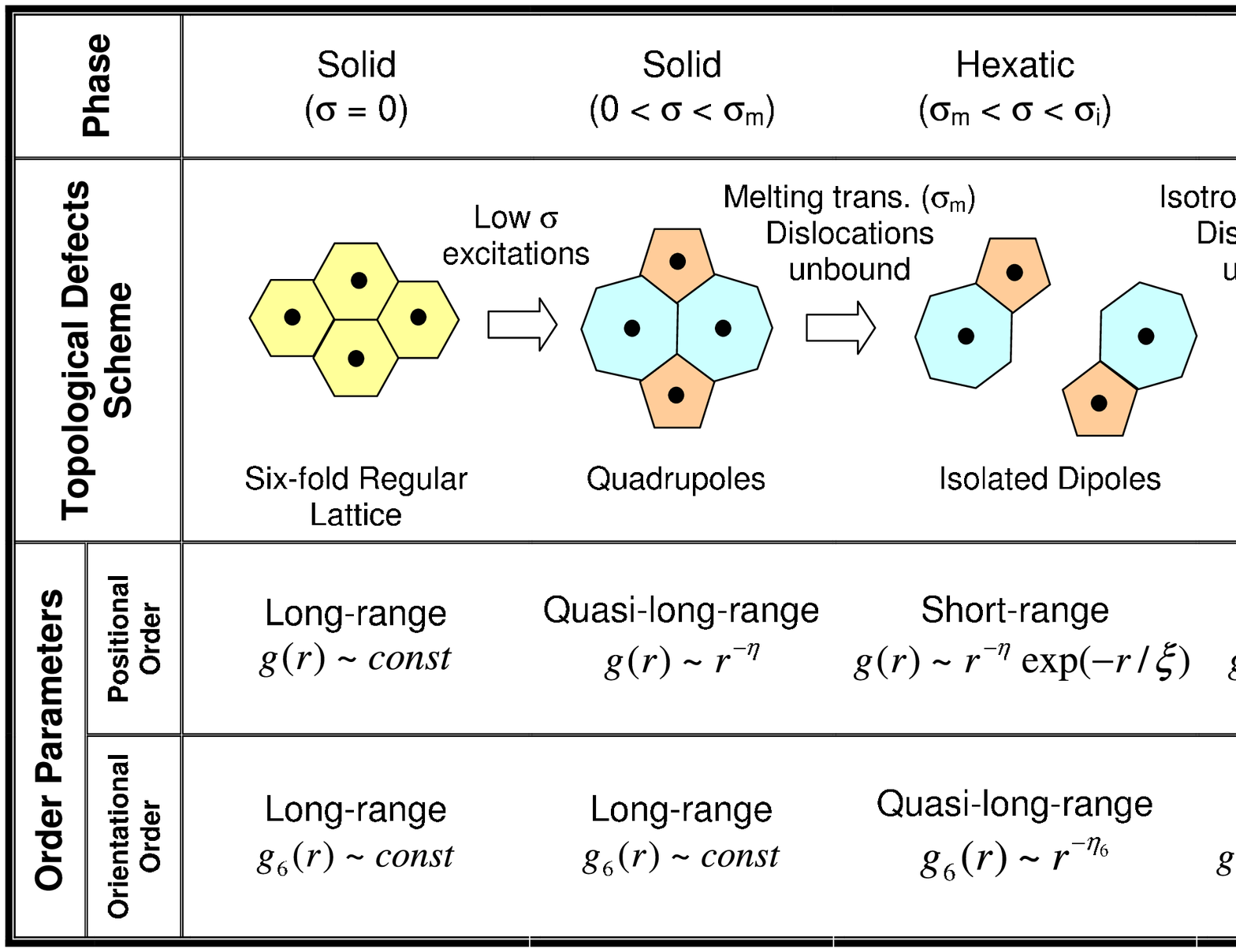,width=\columnwidth,angle=0}
\end{center}
\caption{Schematic presentation of the main predictions 
of the KTHNY theory for two dimensional melting. See the main text for
more details.}
\label{KTHNY}
\end{figure}

\begin{itemize}

\item Only at zero temperature the lattice ordering can be perfect
  while, for any non-vanishing temperature, defects appear. Below a
  first critical point (denoted $\sigma_m$ here) defects are tight
  together in quadrupoles, hindering translational order. As a result,
  translational correlation functions decay algebraically in space
  (with continuously varying exponents), as corresponds to {\it quasi
    long-range order}.  Orientational correlations (see below for a
  precise definition) decay at long distances to a non-vanishing
  constant value, corresponding to true long-range orientational order
  \cite{explanation}.

\item Above a first critical point, $\sigma_m$, dipoles/dislocations
  unbound from quadrupoles, destroy translational order (i.e.
  translational correlations decay exponentially fast). They also
  affect orientational correlations which decay algebraically with a
  continuously variant exponent $\eta_6$ obeying $\frac{1}{4}\leq
  \eta_6 \leq\frac{1}{3}$ and a diverging correlation length, i.e.
  they exhibit quasi-long-range orientational order. This is the, so
  called, {\it hexatic} phase.

\item Above a second (isotropic) critical point, $\sigma_i$,
  monopoles/disclinations unbind from dipoles, hindering quasi long
  range orientational order. Both translational and orientational
  correlations decay exponentially. The associated correlation lengths
  diverge as stretched exponentials of the form $a \exp(b
  \Delta^{-1/2})$ where $\Delta$ is the distance to $\sigma_i$.  This
  is the isotropic (also called disordered or liquid) phase.
\end{itemize}

This scenario has been verified in a number of numerical
\cite{Hexatic,Strandburg} and experimental
\cite{Hexatic-exp,Strandburg} studies, while it was {\it not} verified
in others \cite{JJ,QG}.

Despite of its success, the KTHNY is not the only plausible theory of
two-dimensional melting. A competing one was proposed by Chui
\cite{Chui}, who argued that some systems should exhibit a unique
first-order melting transition mediated by the appearance of ``grain
boundaries''. In this picture, chains of defects appear limiting
ordered grains and separate neighboring mismatching domains. The main
difference between this theory and the KTHNY one is that here the
transition appears owing to a {\it collective excitation} of defects.
In some systems, it has been shown that the melting transition can
change from KTHNY to first order upon changing some parameter
\cite{Strandburg}, so both theories can be taken as complementary.

\section{Analysis of the nonequilibrium melting/freezing transition}

To determine the plausibility of the KTHNY scenario (or, instead, that
of competing theories) for the nonequilibrium transition under study,
we analyze in this section translational and orientational order in
numerical simulations of Eq.(\ref{langevin}). 

\subsection{Translational Order}

Translational order can be studied by measuring the two-point radial
correlation function:
\begin{equation}
  g(r=\left|\textbf{r}-\textbf{r}'\right|) \propto \left\langle 
    \rho(\textbf{r}) \rho(\textbf{r}')\right\rangle -\left\langle
    \rho(\textbf{r})\right\rangle \left\langle
    \rho(\textbf{r}')\right\rangle, 
\end{equation}
where $\left\langle \cdot \right\rangle$ stands for averages in the stationary
state taken over all pairs of particles at generic positions $\textbf{r}$ and
$\textbf{r}'$ separated by a distance $r=\left|\textbf{r}-\textbf{r}'\right|$.
\begin{figure}
\begin{center}
\epsfig{figure=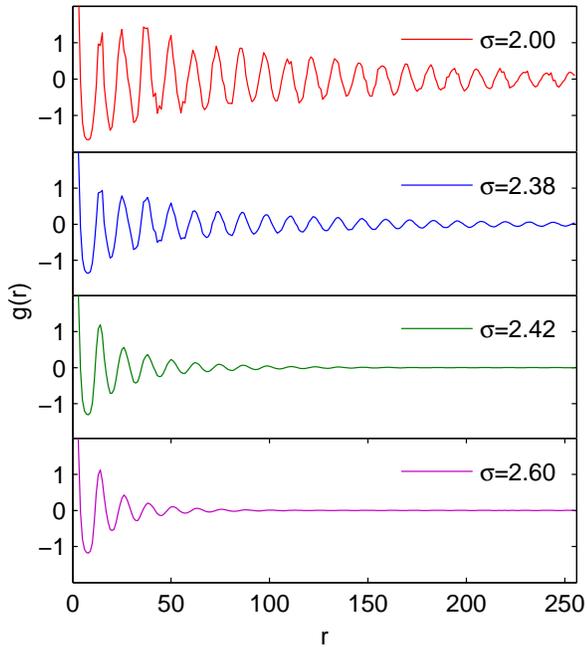,width=\columnwidth,angle=0}
\end{center}
\caption{Radial correlation function, $g(r)$ as a function of $r$ for
  different values of $\sigma$ and linear size $L=512$.}
\label{gr}
\end{figure}

\begin{figure}
\begin{center}
\epsfig{figure=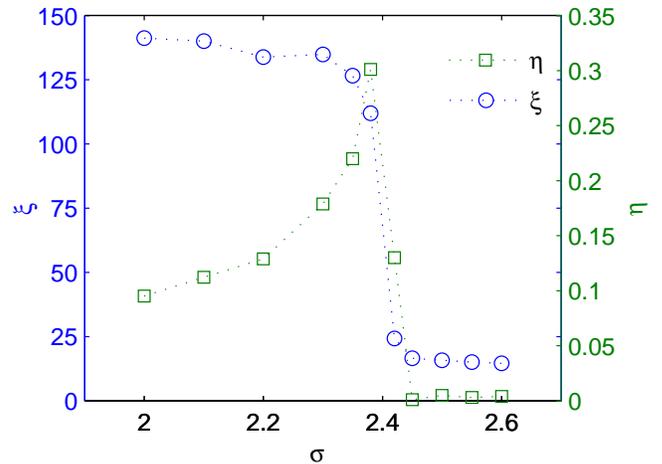,width=\columnwidth,angle=0}
\end{center}
\caption{Correlation length, $\xi$ and exponent $\eta$, obtained by
  fitting $g(r)$ in Fig.\ref{gr} to Eq.(\ref{fit}).  Note the abrupt
  increase of the correlation length around $\sigma =2.39$ from a
  small value above to a large saturating value (roughly equal to $L/4
  =128$, i.e. half of the maximum radius). The exponent $\eta$ changes
  continuously, in remarkable agreement with the KTHNY predictions.}
\label{gr2}
\end{figure}

In Fig.\ref{gr} we plot $g(r)$ as a function of $r$ for different
values of $\sigma$ above and below the critical point.  For all
parameter values the wavy curves reveal the clustered nature of the
density distribution. To determine their asymptotic trends, we analyze
the envelope of such curves, and fit it to the behavior predicted by
the KTHNY theory:
\begin{equation}
g(r) \propto r^{- \eta} e^{-r/\xi} 
\label{fit}
\end{equation}
where $\eta$ and $\xi$ are fitting parameters.  Note that, the
power-law corresponds to the KTHNY prediction while the exponential
allows to describe finite-size induced cut-offs.  Fig.\ref{gr2} shows
the results of such a fit for different values of $\sigma$. In all
cases, the fit correlation coefficient is large than $0.92$. Note
first the abrupt jump of the translational correlation length $\xi$ at
the critical value $\sigma \approx 2.39$; in the ordered phases it
converges to a saturation value controlled by system size (see figure
caption), while in the disordered one it takes much smaller values. On
the other hand, the exponent $\eta$ changes continuously from a value
nearby $0.31(1)$ at the critical point (in perfect agreement with the
KTHNY prediction, which imposes $1/4 < \eta(\sigma_m) < 1/3$) to
smaller values as $\sigma$ is decreased, confirming the generic
algebraic decay of $g(r)$ in the solid phase, with an exponential
cut-off given by the system size.

To check the internal consistency of our results, we also estimate
$\eta$ by analyzing the previously obtained structure-function
results.  As $S({\bf k})$ is trivially related to $g({\bf r}))$
through a Fourier transform, and $g({\bf r})$ (which depends only on
the modulus of ${\bf r}$, $r$) can be modeled in the solid phase by
$g(r)\sim r^{-\eta} \cos(k_0 r)$ (for simplicity, we omit here the
exponential, system size induced, cut-off $e^{-r/\xi}$) one can find
after simple algebra, that for a two-dimensional system of size $L$
\begin{equation}
S(k,L)= 2 \pi \int_{0}^{L} r^{1-\eta} \cos(k_0 r) J_0(k r)dr,
\label{eq:slk}
\end{equation}
where $J_0$ is the zero-order Bessel function. Using the asymptotic
behavior of $J_0(k r)$ it is not difficult to obtain that
\begin{equation}
S(k_0,L) \sim L^{3/2-\eta}
\label{eq:scalingsk}
\end{equation}
for the mode $k_0$ (which is the only one for which there are not
destructive interferences). 

Actually, in the inset of Fig.(\ref{Structure}) we showed that while
the curves for $S(k_0,L)$ overlap for different system sizes above the
critical point, bellow it, they growth with system size.  Indeed they
grow algebraically with a slightly $\sigma$-dependent exponent which
for all $\sigma$ values is in the interval $[1.24,1.36]$.  Therefore,
exploiting Eq.(\ref{eq:scalingsk}) we derive a value for $\eta$ which
is in the interval $[0.14,0.26]$ in rather good agreement with the
direct measurements above. Summing up, we have deduced, in two
independent ways, results consistent with algebraically decaying
correlations as predicted by the KTHNY scenario.


\subsection{Orientational Order}

To quantify the degree of orientational order of a given configuration
in the steady state, we first identify the clusters using the
algorithm described above. After that, a Voronoi tessellation of the
system configuration is constructed \cite{Voronoi}. For each
configuration, the corresponding tessellation gives as output the list
of clusters and the set of nearest neighbors of each. Finally, for
each cluster $j$, we define \cite{Strandburg,KTHNY,Nelson}
\begin{equation}
\psi_{6}(j)=\frac{1}{N_j}\sum_{k=1}^{N_j} e^{i6\theta_{jk}},
\end{equation}
where the sum extends over the $N_j$ neighbors of cluster $j$;
$\theta_{jk}$ is the angle between the centers of mass of clusters $j$
and $k$ and an arbitrary fixed reference axis.
 The average of $\psi_{6}$ over different clusters,
\begin{equation}
\psi_6=\left|\frac{1}{N_c}\sum_{j=1}^{N_c} \psi_{6}(j)\right|,	
\end{equation}
where $N_c$ is the total number of clusters, is a global {\it orientational
  order parameter}.  An associated susceptibility can be also defined,
$\chi_6=N_c\left(\left\langle 
\psi_6^2\right\rangle-\left\langle \psi_6\right\rangle^2\right).$

\begin{figure}
\begin{center}
\epsfig{figure=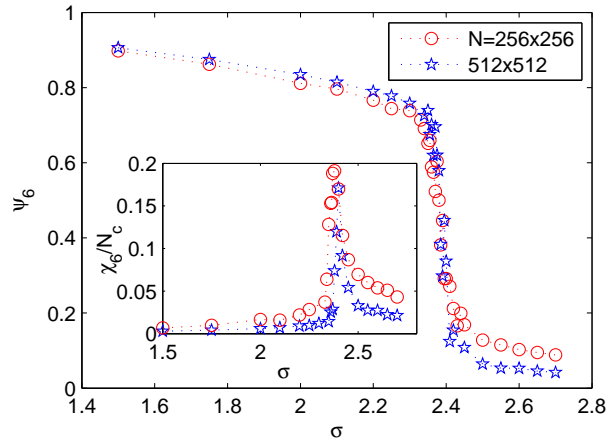,width=\columnwidth,angle=0}
\end{center}
\caption{
Main plot: Stationary orientational order parameter, $\psi_6$, as a
function of $\sigma$ for two different system sizes, $L=256$ and
$512$. Inset: Orientational susceptibility, $\chi_6$, as a function of
$\sigma$ for $L=256$ and $L=512$.}
\label{Orientational}
\end{figure}

Fig. \ref{Orientational} (main plot) shows $\psi_6$ as a function of
$\sigma$ for two different system sizes. For $L=512$, $\psi_6$ changes
abruptly in the interval from a large value for $\sigma$ below
$\approx 2.38$ to relatively small ones above $2.40$. Note that finite
size effects operate in opposite directions below and above the
transition and the curves intersect at some point between these two
regimes (resembling what happens for the Binder cumulant at continuous
phase transitions). This provides a useful criterion to locate the
critical point; using the two available sizes, the best estimation is
$\sigma=2.39(2)$, and strongly suggests that the transition is
continuous in agreement with the KTHNY scenario. On the other hand,
the susceptibility (inset of Fig.  \ref{Orientational}) exhibits a
sharp peak, which moves slightly to the right with increasing system
size, being located at $\sigma=2.39(1)$ for $L=512$.

\begin{figure}
\begin{center}
\epsfig{figure=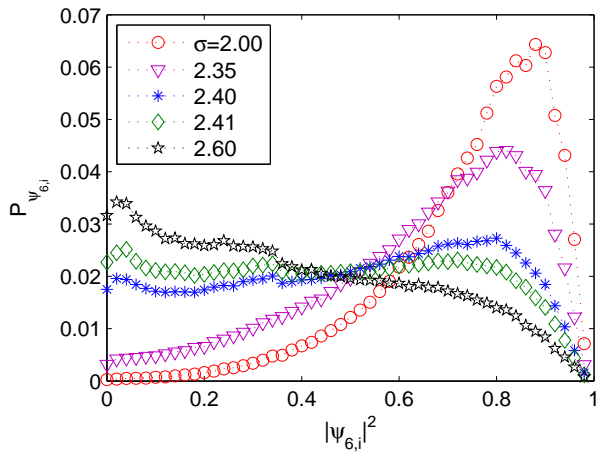,width=\columnwidth,angle=0}
\end{center}
\caption{Probability distribution functions of $\left|\psi_{6}(i)\right|^2$
for different values of $\sigma$ and linear size $512$.  }
\label{psi6pdf}
\end{figure}

The probability distribution function (pdf) of $\psi_{6}(j)$ provides
additional information about the nature of the transition. In Fig.
\ref{psi6pdf} we plot the pdf as a function of
$\left|\psi_{6}(j)\right|^2$ (to have only positive values) for
various $\sigma$ and size $L=512$. At small values of $\sigma$ (as
$2.00$ or $2.35$) the pdf is unimodal and peaked around a high value.
Increasing the noise strength the peak becomes less pronounced.  For
noise values around the transition point, $\sigma \in
\left[2.40,2.42\right]$ the pdfs are rather flat, a new peak appears
nearby zero, and the average value shift (in an apparently continuous
way) from a high value in the ordered phase to a small one in the
disordered one.

\begin{figure}
\begin{center}
  \epsfig{figure=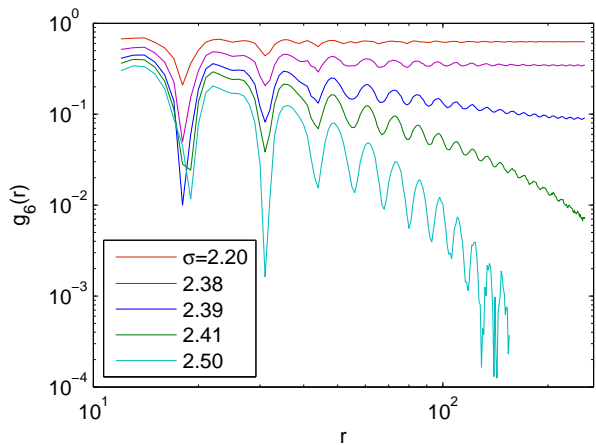,width=\columnwidth,angle=0}
\end{center}
\caption{Log-log plot of the (spherically averaged) orientational correlation
  function, $g_6(r)$,  for different values of $\sigma$ and linear size $512$.}
\label{g6}
\end{figure}

To further check the predictions of the KTHNY theory it is necessary
to determine the two-point orientational correlation function:
\begin{equation}
g_6(r)=\left\langle
\psi_6(j)
\psi_6(k)\right\rangle,
\end{equation}   
where the average is taken over all pairs $j,k$ of clusters separated by a
distance $r$ (results shown in Fig. \ref{g6}).
While for $\sigma >2.40$ the envelope of the curves veers down in a
double-logarithmic plot (i.e. decay exponentially) for values smaller
than $\sigma=2.39$ the envelopes converge to constant values in the
long distance limit, signaling the emergence of true long-range
orientational order in the crystal-like phase. We find no evidence of
a hexatic phase, characterized by generic power-law decaying
orientational correlations. If it actually exists it should be very
narrow and could not be detected within our present numerical
resolution.  Indeed, we have scanned $\sigma$ in steps of $0.005$
units (not shown), and we always find that the envelopes either bend
down below certain separatrix (i.e. the critical point, which here we
estimate to be at $\sigma=2.395(5)$) or flatten above such a value.

\subsection{Analysis of the isotropic phase}

As synthesized above, the KTHNY theory predicts that in the disordered or
isotropic phase \cite{KTHNY,Strandburg,Nelson}
\begin{equation}
g_6(r) \sim r^{\eta_6} e^{-r/\xi_6},
\label{g6liquid}
\end{equation}
where $\eta_6=1/4$ is a critical exponent and $\xi_6$, the {\it orientational
  correlation length}. The theory also predicts a divergence in $\xi_6$ as the
critical point is approached
\begin{equation}
  \xi_6(\Delta)=a_{\xi} e^{b_{\xi}\Delta^{-1/2}},
\label{stretched}
\end{equation}
where $\Delta$ is the distance to the critical point and $ a_{\xi}$ and
$b_{\xi}$ are constants. A similar, stretched exponential behavior is also
predicted for the susceptibility,
$\chi_6(\Delta)=a_{\chi}e^{b_{\chi}\Delta^{-1/2}}$.

Both $\xi_6$ and $\eta_6$ can be measured by fitting the envelopes of
$g_6(r)$ for different values of $\sigma$ to Eq.(\ref{g6liquid}). We
have performed a two-parameter ($\xi_6$ and $\eta_6$) fit, and obtain
the results summarized in Table \ref{fit1}. Observe the very fast grow
of $\xi_6$ upon approaching the transition point. The estimations of
$\eta_6$ are very close to the KTHNY value $\eta_6=1/4$ (actually,
they become indistinguishable by including error-bars). Not
surprisingly, close to the critical point, the correlation
coefficient, $Corr$, is worse than far from it.
\begin{table}
	\centering
	\begin{tabular}{|c|c|c|c|}
          \hline
          $\sigma$ & $\xi_6$ & $\eta_6$ & $Corr$ \\
          \hline
          \hline
          2.39 & 291.8 & 0.320 & 0.977 \\
          2.41 & 70.5 & 0.298 & 0.997 \\
          2.45 & 47.8 & 0.299 & 0.995 \\
          2.50 & 30.6 & 0.261 & 0.998 \\
          2.55 & 24.1 & 0.245 & 0.997 \\
          2.60 & 22.5 & 0.254 & 0.996 \\
          \hline
	\end{tabular}		
	\caption{Correlation length $\xi_6$ and critical exponent
          $\eta_6$ at different values of $\sigma$ in the isotropic
          phase. Results obtained by fitting the (envelopes of the)
          curves for $g_6(r)$ to Eq.(\ref{g6liquid}). The last
          column shows the correlation coefficient, $Corr$ between the
          numerical data and their fits.} 
        \label{fit1}
\end{table}

\begin{figure}
\begin{center}
\epsfig{figure=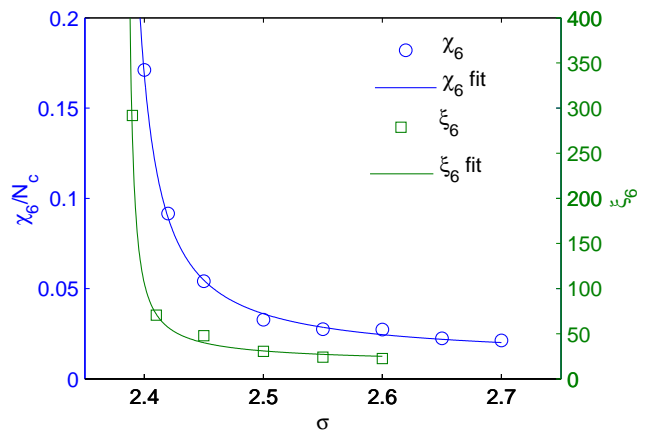,width=\columnwidth,angle=0}
\end{center}
\caption{Orientational susceptibility, $\chi_6$, and correlation length,
  $\xi_6$, in the isotropic phase. The solid lines represent the best fit to
  the stretched exponential function, Eq.(\ref{stretched}).}
\label{figfit}
\end{figure}

In Fig. \ref{figfit} we show the results obtained by fitting the
simulation results to the predicted stretched exponential behavior
Eq.(\ref{stretched}).  The fit is excellent in both cases and allows
for estimations of the critical point location: $2.38(2)$ using
$\xi_6$ and $2.37(2)$ using $\chi_6$.

These results provide a strong backing for a KTHNY scenario in the
isotropic phase.

\subsection{Summary of the numerical observations}

Summing up, {\it the numerical analyses detailed above are consistent
  with the KTHNY theory in any respect, except for the fact that no
  evidence of a hexatic phase is found}. Our results are compatible
with a scenario very similar to the KTHNY one, but with the two
transition points merging into a unique one, occurring somewhere
between $\sigma =2.38$ and $\sigma=2.39$.  No evidence is found of a
hexatic phase. A scenario analogous to this, i.e. a one-stage
continuous melting transition, has been found in other equilibrium and
nonequilibrium systems \cite{JJ,QG}.

\begin{figure*}[t]
\begin{center}
\epsfig{figure=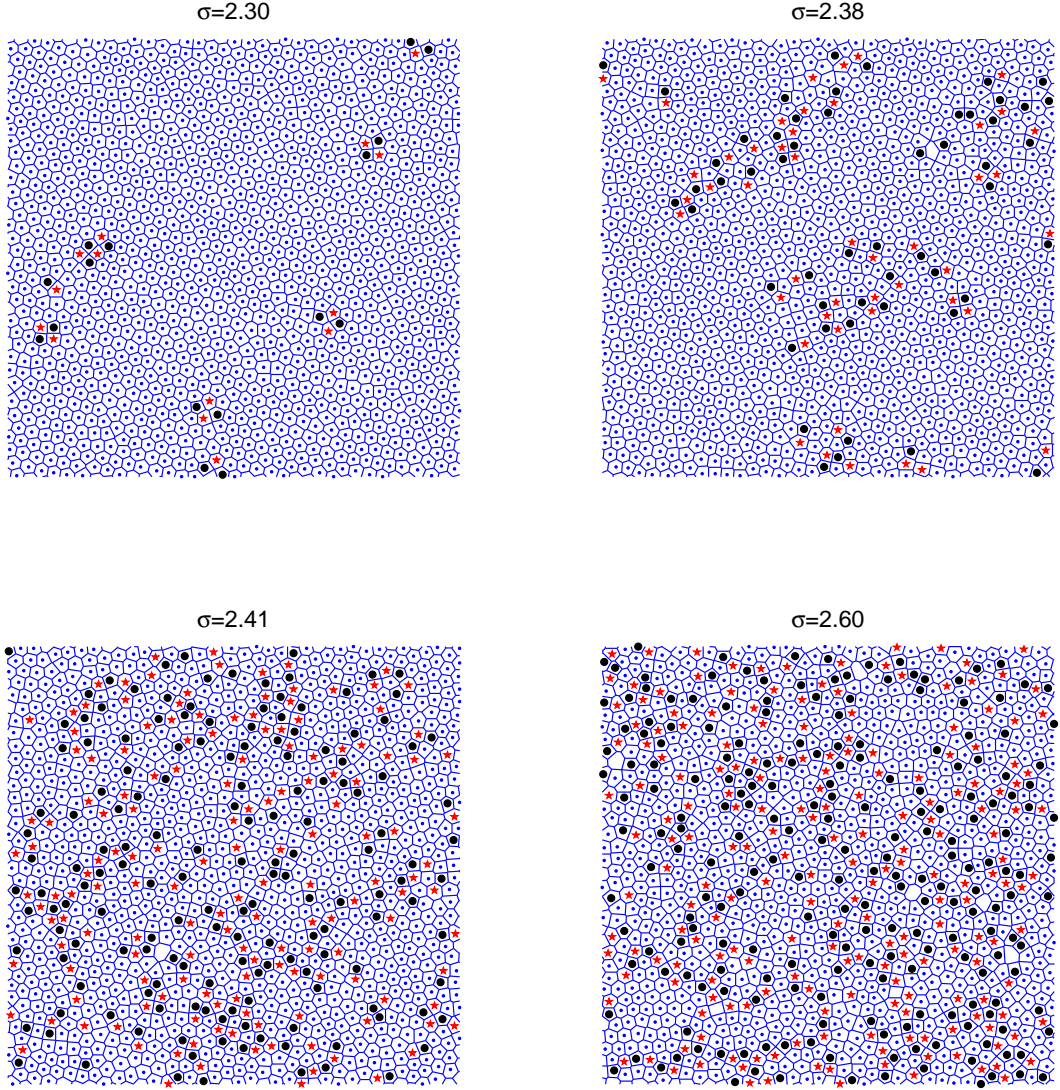,width=\textwidth,angle=0}
\end{center}
\caption{Defect maps at $4$ different values of the noise amplitude,
  $\sigma$ and size $L=512$. Lines define a Voronoi tessellation;
  five-fold, six-fold, and seven-fold clusters are marked with black
  circles, central dots, and red stars respectively. The (few) empty
  polygons correspond to $4$-fold, $8$-fold, and $9$-fold defects. (i)
  Upper left panel (low noise intensity; $\sigma=2.30$); only
  localized quadrupoles appear.  (ii) Upper right panel ($\sigma=2.38$
  nearby the critical point); a proliferation of isolated
  dipoles/dislocations can be observed and a small number of isolated
  monopoles/disclinations can also be observed. (iii) Lower left panel
  ($\sigma=2.41$ slightly above the critical point); although isolated
  dipoles and monopoles exist, there is a clear tendency to form
  defect condensates with string-like geometry. (iv) Lower right panel
  ($\sigma=2.60$ well above the transition point); defects proliferate
  and the tendency to form string-like structures is maintained.}
\label{Defectmaps}
\end{figure*}

\section{Defect analysis}
Inspired by the analyses in \cite{QG}, in this section we check the KTHNY
assumptions rather than its predictions. As briefly explained above, the
theory assumes that the number of dipoles/dislocations
(monopoles/disclinations) throughout the first (second) transition is small
and that they are generated progressively in a smooth way as the temperature
is risen. This allows to treat the first (second) transition as a weakly
interacting gas of dipoles/dislocations (monopoles/disclinations).

To check if this picture describes properly the behavior of our
model, we scrutinize how defects appear and proliferate upon rising
the noise amplitude. In Fig.  \ref{Defectmaps} we plot defect maps for
$4$ different values of $\sigma$.  The lines correspond to the Voronoi
construction for a given configuration; five-fold, six-fold, and
seven-fold clusters are marked with black circles, central dots, and
red stars respectively (higher and lower order defects correspond to
blank clusters). As the noise amplitude is increased the total number
of defects grows. While below the critical point only quadrupoles are
significatively present, nearby the transition point isolated dipoles
start unbinding.  Even if unbound, they have a clear tendency to bunch
together, and indeed, at slightly larger noise-amplitudes, in the
isotropic phase, defects bunch together showing a tendency to form
string-like structures. Deep into the isotropic phase defects
proliferate, and extend through the system keeping, in any case, the
propensity to form condensed string-like structures.

In this respect, it is noteworthy that Fisher, {\it et
  al.} 
\cite{Fisher} predicted, within the KTHNY theory, that monopoles can
appear at correlated locations showing a strong tendency to arrange
themselves in small-angle grain boundaries.  Contrarily, the
condensates we find seem to be dominated by strings of dipoles.

\begin{figure}
\begin{center}
\epsfig{figure=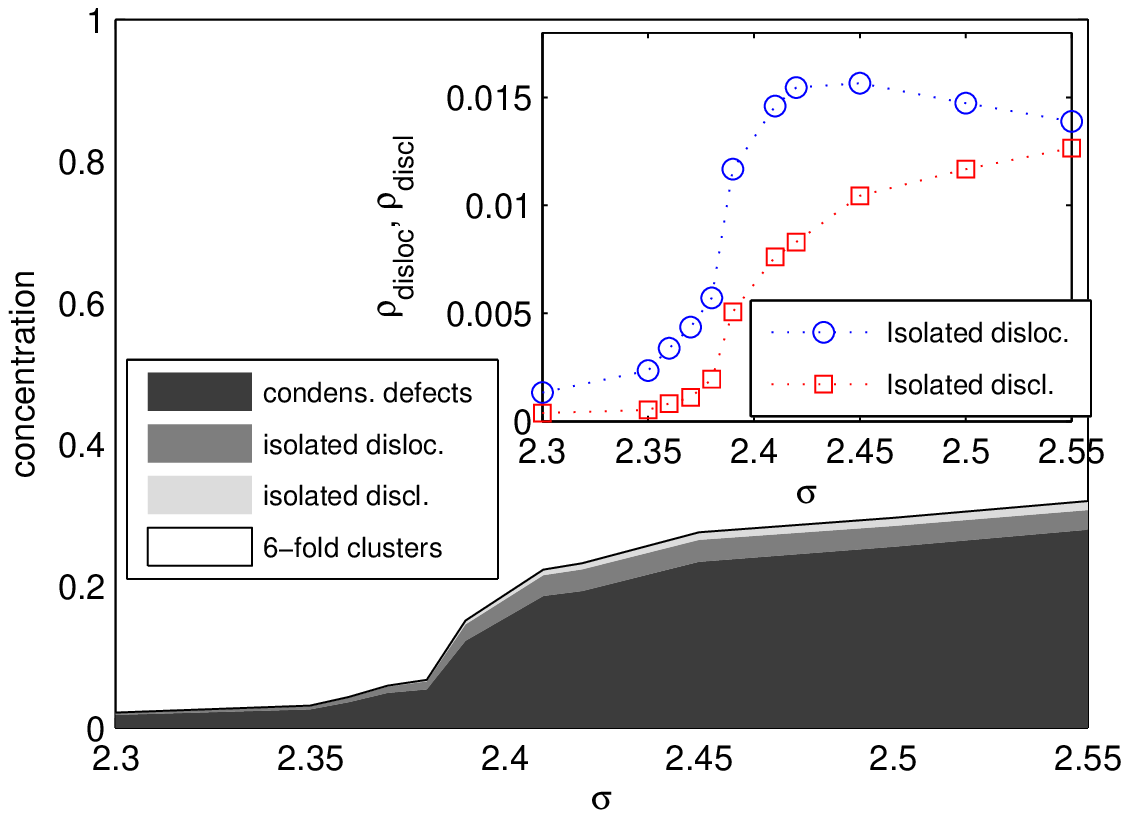,width=\columnwidth,angle=0}
\end{center}
\caption{Main: Accumulated density of defects of the various possible types as
  a function of the noise amplitude, $\sigma$.  Inset: Relative isolated
  dislocation and disclination concentrations versus $\sigma$. }
\label{Defectdensity1}
\end{figure}

To quantify the phenomenology above, we measure the density of defects
(quadrupoles, dipoles and monopoles) as a function of noise amplitude
(see Fig.\ref{Defectdensity1}). The density of defects (of any type)
grows from a small value below $\sigma=2.3$ to a large value in the
isotropic phase. The increase is very steep around $\sigma=2.38$ for
both dipole and monopole densities, suggesting the existence of a
unique unbinding transition.

To complement the previous plot, the inset of Fig.
\ref{Defectdensity1} shows the density of isolated
dipoles/dislocations and monopoles/disclinations versus $\sigma$,
confirming the previous observations, that the steepest increase
occurs at $\sigma \approx 2.39$.  Note that the total number of
monopoles grows continuously upon increasing $\sigma$, while the
presence of a relative maximum in the density of dipoles reflects the
competition between two conflicting tendencies: the unbinding of
dipoles from quadrupoles and the liberation of monopoles (and
formation of condensed defects) from free dipoles.

\begin{figure}
\begin{center}
\epsfig{figure=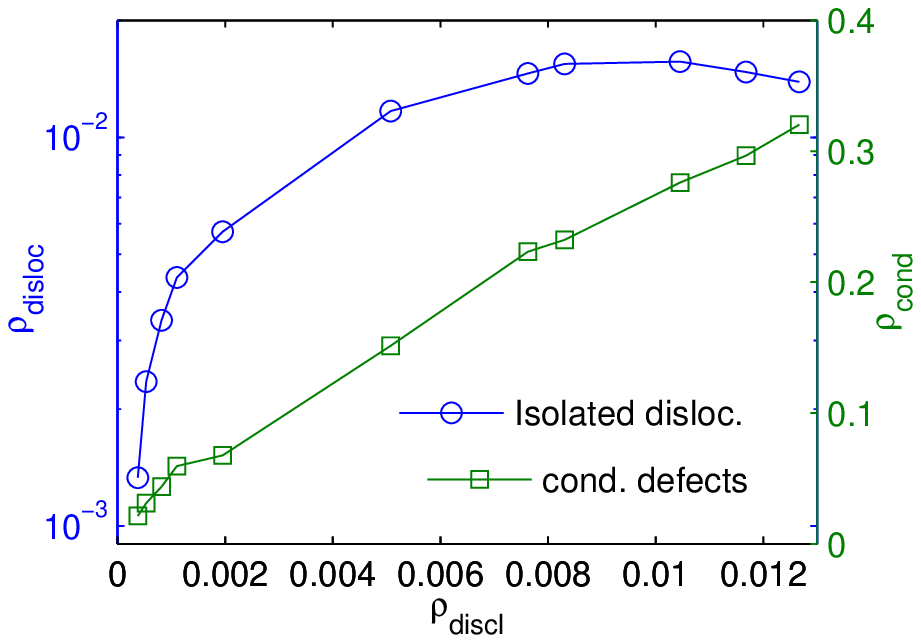,width=\columnwidth,angle=0}
\end{center}
\caption{
  Dislocation density (left axis and circles) and total defect density (right
  axis and boxes) in logarithmic scale, as a function of the isolated
  disclination density ($L=512$).}
\label{Defectfinal}
\end{figure}

Finally, Fig. \ref{Defectfinal} shows the densities of isolated
dipoles and of condensates as a function of the density of monopoles.
This plot reveals a strong correlation between monopoles and dipoles
and, more importantly, it illustrates how they both vanish at roughly
the same point, suggesting again (up to finite size effects) the
existence of a unique one-stage transition.  Indeed, if the dipoles
unbinding occurred before the monopoles one, the line joining the
circles should intersect the vertical axis which is not the case. It
is also at such a unique transition point that condensed defects
appear, being their density (roughly) linearly correlated with the
density of monopoles.

In summary, a detailed analysis of defects in Voronoi tessellations
supports the interpretation that both dipoles and monopoles unbind at
a unique critical point. At such a transition point condensed defects
are generated in an apparently continuous way.  In a future
publication we will analyze in a more detail the correlations between
isolated monopoles, dipoles and condensates to further clarify the
analogies and differences with the standard KTHNY scenario, and try to
develop a theoretical framework for one-stage continuous melting
transitions.

\vspace{0.25cm}
\section{Discussion and Conclusions}

Motivated by the observation of remarkably regular arrays of clusters
formed by bacteria growing in Petri dishes and related problems, we
have revisited the {\it individual-based interacting Brownian bug} (IBB)
model in which birth rates are local-density dependent
\cite{bugs1,bugs2}. Apart from an absorbing phase transition, this
model shows a transition from a disordered active phase, in which
particles aggregate in localized clusters, to an ordered or
crystallized active phase, in which clusters self-organize forming
hexagonal arrays. For the sake of generality and aimed at facilitating
analytical studies, rather than studying the discrete model itself we
have analyzed its equivalent continuous Langevin representation. This
stochastic equation, which involves a non-local saturation term as
well as a square-root multiplicative noise, can be numerically
integrated by discretizing it and employing a recently introduced
efficient integration scheme specifically designed to deal with
square-root multiplicative noise \cite{Dornic}.

The main results we obtain are:

(i) First, we have shown explicitly that the continuous model
(truncated to include only the leading terms) reproduces the
phenomenology of the original discrete model, including an absorbing
phase in which the stationary particle density is zero, a disordered
active phase in which the density field is localized in clusters of
activity surrounded by empty regions and, finally an ordered phase
with hexagonal patterns.

(ii) We have studied the ordering transition by analyzing cluster
properties (average velocity, mean-displacement, etc.) as well as by
means of analyses of the structure-function. These studies reveal that
a melting/freezing transition indeed occurs at some value of the noise
amplitude: for small noises, clusters are trapped into hexagonal
configurations as a result of a collective effect, while for
noise amplitudes above threshold they have much more mobility.

(iii) To better understand and quantify the transition we have studied
both translational and orientational correlation functions.  In
particular, we have verified that, in the solid phase, translational
correlation functions exhibit generic power-law decay even if with
system-size induced cut-offs. The value of the decay exponent at the
melting transition being in excellent agreement with the KTHNY
prediction.  However, we do not found any evidence of an hexatic
phase, contrarily to what predicted by the standard
(Kosterlitz-Thouless-Halperin-Nelson-Young (KTHNY)) theory.  This same
conclusion is also borne out by analysis of the global orientational
order-parameter. Up to the numerical resolution limit, this suggests
the existence of a unique continuous transition point.

(iv) Above such a transition point, i.e. in the disordered or
isotropic phase, the correlation functions have been shown to decay as
stretched exponentials, in excellent agreement with the KTHNY
predictions.

(v) We have performed a detailed analysis of defects and found that
both free dipoles/dislocations and monopoles/disclinations seem to
unbind from quadrupoles at a unique transition point, above which also
defect condensates are formed.  This is in contrast with the KTHNY
scenario and leads to a continuous one-stage melting transition.

(vi) The unique transition is continuous, so it cannot be explained by
the main alternative theory to KTHNY \cite{Chui} which predicts a
first-order melting transition.

Summing up, the phenomenology of this nonequilibrium model can be only
partially described by the KTHNY theory. While the melting/freezing transition
is indeed characterized by the smooth unbinding of defects this occurs through
a unique continuous transition. The reason for this seems to yield in the fact
that once dipoles/dislocations unbind, the perturbation they generate around
them is large enough as to unbind also monopoles/disclinations: dipoles,
monopoles and the string-like condensates they form are {\it strongly
correlated}.

Note that, the nonequilibrium microscopic dynamics of the interacting
Brownian bug model (or its equivalent Langevin representation) is
responsible for the generation of mesoscopic clusters. Once such
clusters are generated, they interact in an effective way, and the
physics at a macroscopic scale does not seem to differ in any
essential way from other equilibrium problems \cite{JJ,QG} for which a
similar one-stage continuous melting transition has been reported. We
believe that this scenario is not specific of nonequilibrium systems
but is determined by the way in which defects interact among
themselves. This might not depend on the equilibrium versus
nonequilibrium nature of the process but rather on other structural
details influencing the way defects interact among themselves. A more
detailed analysis of defects and defect-correlation will be
investigated in a future work, where we will try to develop a
theoretical framework for one-stage continuous melting transitions.

Let us also emphasize that the models we have analyzed are not the
best choice to explore with high numerical resolution the possibility
of one-stage melting from a general perspective. Effective models,with
a dynamics at the level of clusters (as opposed to having a
microscopic dynamics for particles) would be a much better option from
the computational point of view.

In summary, despite of interesting and not fully understood
differences, the striking patterns produced by the biologically
inspired Langevin equation (\ref{langevin}) resemble very much the
melting/freezing solid/liquid transition in equilibrium systems.

It is our hope that this paper will motivate further studies of (i)
the effect of fluctuations on self-organized nonequilibrium patterns
and (ii) the analogies and differences between the defect-mediated
type of melting transition described here and standard equilibrium
melting scenarios.

\vspace{0.5cm} {\bf ACKNOWLEDGMENTS} We acknowledge useful comments and
discussions with F. de los Santos. This work was supported in part by
the Spanish projects FIS2005-00791 and FIS2004-00953 (Ministerio de
Educaci\'on y Ciencia), FQM-165 (Junta de Andaluc\'ia), and the
European Commission through the NEST-Complexity project PATRES
(043268).


\begin{thebibliography}{99}
\bibitem{nb} Since the only factor impeding unbounded diffusive spreading 
of particles is their finite lifetime which, in its turn, is
proportional to the inverse of the death rate, $1/\beta_0$, the
average linear size of clusters has to be proportional to
$\sqrt{D/\beta_0}$, as can be numerically verified \cite{bugs1,bugs2}.
\bibitem{linear} The linear dependence in the ordered phase can be 
interpreted as a direct transfer of the ``thermal'' energy introduced
by the noise to the kinetic energy of clusters.
\bibitem{explanation} 
Orientational ordering escapes the premises of the MWH theorem.
Indeed, the local hexagonal ordering induces an effective long-range
interaction not covered by the theorem hypotheses.
\end{thebibliography}
\end{document}